# Determination of Minimal Sets of Control Places for Safe Petri Nets

A. DIDEBAN, H. ALLA

*Abstract*— Our objective is to design a controlled system with a simple method for discrete event systems based on Petri nets. It is possible to construct the Petri net model of a system and the specification separately. By synchronous composition of both models, the desired functioning closed loop model is deduced. Often uncontrollable transitions lead to forbidden states. The problem of forbidden states is solved using linear constraints. A set of linear constraints allows forbidding the reachability of these states. Generally, the number of these so-called forbidden states and consequently the number of constraints are large and lead to a great number of control places. A systematic method to reduce the size and the number of constraints for safe Petri Nets is given. By using a method based on the Petri nets invariants, maximal permissive controllers are determined. The size of the controller is close to the size of the specified model, and it can be implemented on a PLC in a structural way.

## I. INTRODUCTION

Supervisory control theory is essentially a theory for restricting the behavior of the plant to satisfy given specifications that describe which evolutions of the plant should not be allowed. The theory of Ramadge and Wonham [1][2] is based on the modeling of the systems using formal languages and finite automata. However, the great number of states representing the behavior of system, and the lack of structure in the models, limit the possibility of developing an effective algorithm for the analysis and the synthesis of real systems. To solve these problems, several methods of controller synthesis based on Petri Nets (PNs) were proposed. PNs are a suitable tool to study Discrete Event Systems (DES) due to its capability in modeling and its mathematical properties.

In [3] and [4], the authors use the marking invariants to determine algebraically the incidence matrix of the supervisor PNs model. This method is very simple to be used. However, if some transitions are uncontrollable, it does not give the maximal permissive solution. This technique presents two other disadvantages: 1) it is not always possible to describe the specifications by constraints and, 2) the number of constraints can be very large.

A general objective of the control synthesis is to prevent from forbidden states. These states may be deduced from specifications, they can also be deadlock states. A method to minimize the addition of PN places is proposed in [5], it is based on elementary siphons. There are some drawbacks in this paper. Firstly, one can see that it is based on the computation of minimal siphons and secondly the proposed method is not generally optimal. Another problem is that uncontrollable transitions cannot be considered. In [6] and [7], the authors proposed a method for solving the problems of forbidden states by the theory of regions. The advantage of this method is its generality for non safe PNs. However there are some drawbacks for this method:
- Generally the number of control places is large.
- The computation time for solving the set of integer equations can be very large.

In [8], it is shown that it is possible using the linear constraints to specify forbidden states for safe and conservative PNs. The proposed approach is based on the equivalence between the set of forbidden states and the set of linear constraints, which are deducted from it. Using invariants technique presented in [3], allows building a set of control places, which constitute the optimal controller. However, the number of control places and consequently the number of constraints are large and lead to a large number of control places. In [8], it is also shown that some constraints can be replaced by a single constraint; however there is no systematic method to calculate the simplified constraints in a general case. The problem comes from the linear constraints, which can be simplified taking the PN structural properties into account. In [9], a systematic method has been presented to reduce the number of constraints for safe and **conservative** PNs. The equations deduced from P-invariants property in conservative PNs are used for simplifications.

In this paper, a method is proposed to reduce the number of linear constraints for safe PNs. The expected advantage of this method is its applicability for all kind of safe PN models, and not only for the conservative PNs. The time and memory space for simplification is less than those presented in [9].

In our approach and also in [9], we use the Reachability Graph (RG) as an intermediate step for calculating the controller. Although the complexity of the computation of RG is exponential, this calculation is performed off-line. Moreover, the implemented final controller is a PN model,

A. Dideban is with the GIPSA-Lab, ENSIEG, BP46, 38402 Saint-Martin d'Hères, FRANCE , he is also an assistant professor in Semnan university, Iran (e-mail: abbas.dideban@ lag.enseig.inpg.fr).
H. Alla is with the GIPSA-Lab, France. (e-mail: hassane.alla@inpg.fr).

whose size is very close to the starting model. Generally, few control places are added.

In this paper, an important concept of *over–state* will be defined; it corresponds to a set of markings which have the same property. This idea will help us to build the simplest constraints which forbid a greater number of states. A property for the existence of the maximal permissive controller will be analytically proved. In some very particular cases of non conservative PNs, the optimal solution does not exist. We show that this approach allows highlighting this problem in a simple way.

This paper is organized as follows: In Section 2, the motivation and the fundamental definitions will be presented and illustrated via an example. In Section 3, the idea of passage from forbidden states to the linear constraints will be introduced. The concept of **over-state** will be defined in Section 4. Section 5 constitutes the part of this paper where the fundamental properties for the simplification methods are presented. The calculation of maximal permissive controller will be described in Sections 6 and 7.

## II. PRELIMINARY PRESENTATION

In this paper, it is supposed that the reader is familiar with the PNs basis [10] and the theory of supervisory control [1], [2]. Here, we present only the notations and definitions that will be used later.

A PN is represented by a quadruplet $R = \{P, T, W, M_0\}$ where $P$ is the set of places, $T$ is the set of transitions, $W$ is the incidence matrix and $M_0$ is the initial marking. This PN is assumed to be safe; the marking of each place is a Boolean.

**Definition 1:** The set $\{0,1\}^N$ represents all the Boolean vectors of dimension $N$.
❑

A marking of a safe PN containing $N$ places is a vector of the set $\{0,1\}^N$.

The set of the marked places of a marking $M$ is given by a function support defined as below:

**Definition 2:** The function Support($X$) of a vector $X \in \{0,1\}^N$ is:
Support($X$) = the set of marked places in X.
❑

The support of vector $M_0^T = [1, 0, 1, 0, 0, 1, 0]$ is:
  Support ($M_0$) = $\{P_1 P_3 P_6\}$ or more simply:
  Support ($M_0$) = $P_1 P_3 P_6$

To simplify the writing of the formal expressions, we will use the support of a marking instead of its corresponding vector.

$\mathcal{M}_R$ denotes the set of PN reachable markings. In $\mathcal{M}_R$, two sub sets can be distinguished: the set of authorized states $\mathcal{M}_A$ and the set of forbidden states $\mathcal{M}_F$. The set of forbidden states correspond to two groups: 1) the set of reachable states ($\mathcal{M}_{F'}$) which either do not respect the specifications or are deadlock states and, 2) the set of states such that the occurrence of uncontrollable events leads to states in $\mathcal{M}_{F'}$.

The set of authorized states are the reachable states without the set of forbidden states:

$$\mathcal{M}_A = \mathcal{M}_R \setminus \mathcal{M}_F$$

Among the forbidden states, an important subset is constituted by the border forbidden state denoted as $\mathcal{M}_B$.

**Definition 3:** Let $\mathcal{M}_B$ be the set of border forbidden states:

$$\mathcal{M}_B = \{M_i \in \mathcal{M}_F \mid \exists \sigma \in \Sigma_c \text{ and } \exists M_j \in \mathcal{M}_A, M_j \xrightarrow{\sigma} M_i\}$$

Where $\Sigma_c$ is the set of controllable transitions
❑

We will use the following example in order to illustrate the definitions and the results developed in this paper.

Consider a system composed of two machines $Ma_1$ and $Ma_2$ which can work independently. The starting and the end of jobs on these machines are respectively realized by controllable events $c_1$ and $c_2$, and uncontrollable events $f_1$ and $f_2$. When the machine $Ma_1$ ends its job on a part, it stays available for a new job while machine $Ma_2$ has to transfer its produced part in a buffer before beginning a new job (event $t_2$). The specification imposes a sequence of the events $f_1$ and $t_2$. An elementary process corresponds to the treatment of a part by $Ma_1$ followed by the treatment of a part by $Ma_2$. This production is repeated in a cyclic way. The synchronous composition of the system's model and the specification's model is given in Figure 1.

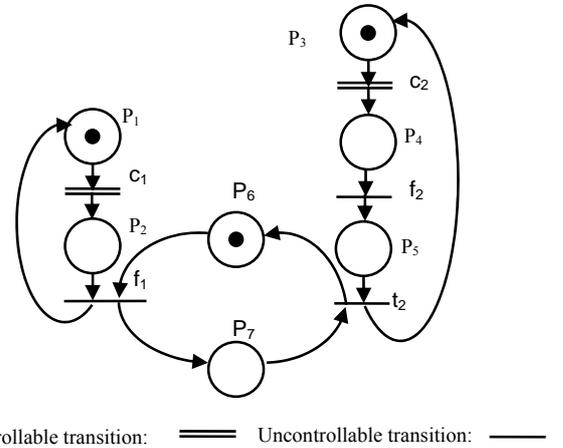

Controllable transition: ══  Uncontrollable transition: ──

Fig. 1 PN model of the system coupled with its specification

The existence of uncontrollable events leads to the existence of forbidden states. The set of forbidden states can be determined by the algorithm established by Kumar [11].

Figure 2 gives the reachability Graph of the PN presented in Figure 1. The forbidden states are indicated in dark gray and the authorized states in white. In this example, there is no deadlock state.

From the set of forbidden states $\mathcal{M}_F = \{M_5, M_6, M_7, M_8, M_9, M_{10}, M_{12}\}$, we can construct the set of border forbidden states

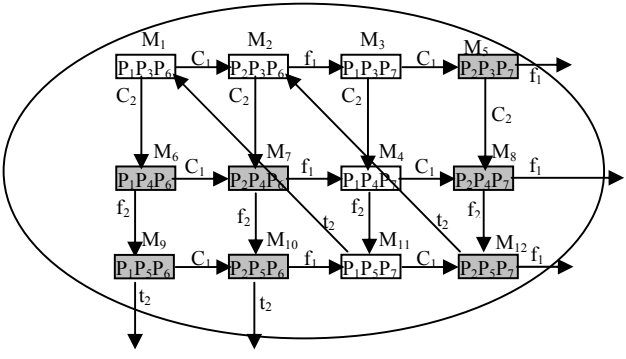

Fig. 2  Reachability graph

$\mathcal{M}_B$:    $\mathcal{M}_B = \{M_5, M_6, M_7, M_8, M_{12}\}$

In a safe PN, as in our example, the inequality $m_1 + m_4 + m_6 \leq 2$ forbids the state $P_1P_4P_6$ [8]. Thus for $N$ forbidden states, we will need $N$ linear constraints. The complexity of the controller model increases when the number of forbidden states increases since we need one control place for each constraint [4]. In this paper, we propose a method to reduce the number and the size of the linear constraints for a given set of forbidden states. To achieve this goal, we need to introduce the important concept of "*over-state*" and some hypothesizes presented below:

**Hypothesis 1:**
  1) All of the events are independent.
  2) The forbidden states are non marked states and all the authorized states are marked.

## III. FROM FORBIDDEN STATES TO LINEAR CONSTRAINTS

Let $M_i$ ($M_i^T = [m_{i1}, m_{i2}, …, m_{iN}]$) be a forbidden state[1] in set $\mathcal{M}_B$ and Support($M_i$) = $\{P_{i1} P_{i2} P_{i3}... P_{in}\}$ be the set of marked places of $M_i$. From a forbidden state, a linear constraint can be constructed [8].

The linear constraint deduced from the forbidden state $M_i$ is given below. The state $M_i$ does not verify this relation. Thus, by applying this relation, $M_i$ will be forbidden.

$$\sum_{k=1}^{n} m_{ik} \leq n - 1$$

Where $n$ = Card [Support ($M_i$)] is the number of marked places of $M_i$, and $m_{ik}$ the marking of place $P_{ik}$ of state $M_i$.

Let $M$ ($M^T = [m_1, m_2,…, m_N]$) be a general marking and $M_i$ be a forbidden state. The constraint (forbidding state $M_i$) is denoted as $c_i$ and can be rewritten in the following form:

$$M_i^T \cdot M \leq \text{Card [Support } (M_i)] - 1$$

For example if:
$M_i^T = [0, 1, 1, 0, 0, 0, 1] \Rightarrow$ Card [Support ($M_i$)] = 3
$$\Rightarrow m_2 + m_3 + m_7 \leq 2 \qquad (1)$$

Verifying relation 1 is equivalent to forbid state $M_i$ when the PN model is conservative. However, in a safe PN not necessarily conservative, this equivalence is not always true. This problem will be discussed later. This equivalence is necessary to obtain the optimal supervisor.

[1] Where there is no ambiguity, the word *border* will be omitted.

## IV. OVER – STATE

### A. Definition of an over – state

The concept of over-state is very important in this paper. An over-state can represent a complete state or a part of this one. In the example of two machines, $P_2P_3P_6$ is a complete state that represents the situation of both machines and the specification. $P_2P_3$ is an over-state of this state that represents a partial state of the system. We have seen that a state can be forbidden by a linear constraint. In the same way, it is possible to forbid an over-state by its corresponding constraint.

**Definition 4:** Let $M_2 = P_{21} P_{22} …P_{2m}$ an accessible state, $M_1 = P_{11} P_{12} …P_{1n}$ will be an over-state of $M_2$ if:   $M_1 \leq M_2$
❏

For example $M_1 = P_1P_3$ is an over-state of $M_2 = P_1P_3P_6$.

The name "*over-state*" is used because the constraint corresponding to an over-state holds the state's constraint. For example the constraint $m_4 + m_6 \leq 1$ corresponding to the over-state $M_1 = P_4P_6$ holds both following constraints:

$$m_1 + m_4 + m_6 \leq 2$$
$$m_2 + m_4 + m_6 \leq 2$$

These two constraints forbid the states $M_6 = P_1P_4P_6$ and $M_7 = P_2P_4P_6$. $P_4P_6$ is an over-state of both states $P_1P_4P_6$ and $P_2P_4P_6$ which could be verified by $M_1 \leq M_6$ and $M_1 \leq M_7$. Thus only using the constraint $m_4 + m_6 \leq 1$ both states $M_6$ and $M_7$ will be forbidden. However, this reduction is not always so simple; sometimes it is possible that the simplified constraint forbids also some authorized states. We present below a method of simplification, which guarantees that the constraints forbid only the forbidden states.

**Remark 1:** With each over-state $b_i$, we associate a constraint $c_i$ in the following way:

$$b_i = (P_{i1}P_{i2}P_{i3}…P_{in}) \Rightarrow c_i = (P_{i1}P_{i2}P_{i3}…P_{in}, n\text{-}1)$$

That means:    $m_{i1} + m_{i2} + … + m_{in} \leq n\text{-}1$
❏

**Remark 2:** There are two relations of inclusion, which work in opposite direction: a set inclusion and a marking inclusion.  Let $M_1 \leq M_2$:
1) The set of the marked places in the over-state $M_1$ is included in the set of the marked places in the state $M_2$.

2) The set of the markings covered by $M_1$ contains those covered by marking $M_2$.
❏

**Property 1:** Let $M_1$ and $M_2$ be two vectors of $\{0, 1\}^N$ and $c_1$ and $c_2$ be two corresponding constraints. If $M_1 \leq M_2$ ($M_1$ is an over-state of $M_2$) and $c_1$ is true, then $c_2$ is also true:

$M_1 \leq M_2$ and $c_1$: $M_1^T . M \leq$ Card[Support($M_1$)]- 1

$\Rightarrow c_2$: $M_2^T . M \leq$ Card[Support($M_2$)]- 1

❑

All the demonstrations of the properties used in this paper can be found in [12].

### B. Set of over-states

We have seen that for forbidding a state it is enough to forbid its over-state but which over-state? This question will be answered in this paper. To achieve this goal, we need to construct the set of over-states for the forbidden states.

Firstly, we calculate the set of over-states for each state and then the union of all the over-states giving the final set.

**Definition 5:** Let $M_i = \{P_{i1} P_{i2} P_{i3} ... P_{in}\}$ be a state of the system. The set of the over-states of $M_i$, denoted as $M_i^{over}$, is equal to the set of the subsets of $M_i$ without the empty set.

❑

For example, the state $M_1 = P_1 P_4 P_6$  gives:

$M_1^{over} = \{P_1, P_4, P_6, P_1P_4, P_1P_6, P_4P_6, P_1P_4P_6\}$

Among, the set of forbidden states in $\mathcal{M}_F$, only the borders states have to be considered in the controller synthesis. Let $\mathcal{M}_B$ be this set and $B_1$ be the set of over-states of $\mathcal{M}_B$.

$$B_1 = \bigcup_{i=1}^{Card(M_B)} M_i^{over}$$

### V. BUILDING THE REDUCED SET OF OVER-STATES

It is needed to build two sets of over-states; a set of the authorized over-states $A_1$, and that of the forbidden states $B_1$. It is obvious that no over-state of $A_1$ must be forbidden. Thus it is necessary to remove from the set $B_1$, all the over-states which are in $A_1$.

**Property 2:** Let $B_1$ be the set of over-states of $\mathcal{M}_B$ and $A_1$ be the set of over-state of $\mathcal{M}_A$ and:

$$B_2 = B_1 \setminus A_1$$

Verifying the set of constraints $C_2$ (equivalent to $B_2$) do not forbid any authorized state.

❑

The proof of this property is obvious.

Another point is that in set $B_2$, it may occur that two over-states $M_1$ and $M_2$ are such as $M_1 \leq M_2$. In that case, $M_2$ is removed giving the set $B_3$ defined formally as follows:

$$B_3 = B_2 - \{M_{2i} \in B_2 \mid \exists M_{2j} \in B_2, M_{2i} \geq M_{2j}\}$$

$B_3$ is the set of the over-states to be forbidden.

For the example of Figure 1, the sets of borders forbidden states and authorized states are:

$M_B = \{P_1P_4P_6, P_2P_4P_6, P_2P_3P_7, P_2P_4P_7, P_2P_5P_7\}$

$M_A = \{P_1P_3P_6, P_2P_3P_6, P_1P_3P_7, P_1P_4P_7, P_1P_5P_7\}$

The different sets $A_1$, $B_1$, $B_2$ and $B_3$ are then calculated below:

$B_1 = M_1^{over} \cup M_2^{over} \cup M_3^{over} \cup M_4^{over} \cup M_5^{over} = \{P_1, P_2, P_3, P_4, P_5, P_6, P_7, P_1P_4, P_1P_6, P_4P_6, P_2P_4, P_2P_6, P_2P_3, P_2P_7, P_3P_7, P_4P_7, P_2P_5, P_5P_7, P_1P_4P_6, P_2P_4P_6, P_2P_3P_7, P_2P_4P_7, P_2P_5P_7\}$

$A_1 = \{P_1, P_2, P_3, P_4, P_5, P_6, P_7, P_1P_3, P_1P_6, P_3P_6, P_2P_3, P_2P_6, P_1P_7, P_3P_7, P_1P_4, P_4P_7, P_1P_5, P_5P_7, P_1P_3P_6, P_2P_3P_6, P_1P_3P_7, P_1P_4P_7, P_1P_5P_7\}$

$B_2 = \{\cancel{P_1}, \cancel{P_2}, \cancel{P_3}, \cancel{P_4}, \cancel{P_5}, \cancel{P_6}, \cancel{P_7}, P_1P_4, P_1P_6, P_4P_6, P_2P_4, P_2P_6, P_2P_3, P_2P_7, P_3P_7, P_4P_7, P_2P_5, P_5P_7, P_1P_4P_6, P_2P_4P_6, P_2P_3P_7, P_2P_4P_7, P_2P_5P_7\}$

$B_3 = \{P_4P_6, P_2P_4, P_2P_7, P_2P_5, P_1\cancel{P_4P_6}, P_2\cancel{P_4P_6}, P_2\cancel{P_3P_7}, P_2\cancel{P_4P_7}, P_2\cancel{P_5P_7}\} = \{P_4P_6, P_2P_4, P_2P_7, P_2P_5\}$

### VI. MAXIMAL PERMISSIVE CONTROLLER

In the previous section, we have determined the set $B_3$, which is the greatest set of over-states that must be forbidden. With the concept of marking covering, we are going to present in the two following sections, the necessary and sufficient conditions to obtain a maximal permissive controller.

With each over-state of $B_3$, we associated a constraint in the following way:

$$b_i = (P_{i1} P_{i2} P_{i3} ... P_{in}) \Leftrightarrow c_i = (P_{i1} P_{i2} P_{i3} ... P_{in}, n\text{-}1)$$

Let $C_3$ be the set of these constraints in our example:

$C_3 = \{(P_4P_6, 1), (P_2P_4, 1), (P_2P_7, 1), (P_2P_5, 1)\}$

This set $C_3$ defines the set of non-forbidden states, denoted as $\mathcal{M}_E$. Now our objective is to compare the set of authorized states $\mathcal{M}_A$ and $\mathcal{M}_E$.

**Remark 3:** Constraint $c_i$ and over-state $b_i$ are equivalent as shown above.

❑

**Definition 7:** Let $B_3 = \{b_1, b_2, ..., b_m\}$ be the set of simplified over-states and $\mathcal{M}_B = \{M_1, M_1, ..., M_N\}$ be the set of border forbidden states. The relation $R: \mathcal{M}_B \times B_3 \rightarrow \{0, 1\}$ is such as:

$$R(M_i, b_j) = \begin{cases} 1 & b_j \leq M_i \ (b_j \text{ is over-state of } M_i) \\ 0 & \text{if not} \end{cases}$$

The covering of a marking is the integer number:

$$Cv(M_i) = \sum_{j=1}^{m} R(M_i, b_j)$$

❑

$Cv(M_i) \geq 1$ mean that forbidden state $M_i$ is covered by at least one over-state.

**Property 3:** The set of non forbidden state $\mathcal{M}_E$ is equal to the set of authorized state $\mathcal{M}_A$ if and only if:

$$\forall M_i \in \mathcal{M}_B \quad Cv(M_i) \geq 1$$

❑

We are going to illustrate the results established above on the example of Figure 1. Property 3 should initially be checked. For this, we construct a table (Figure 3) where the first line represents the set of forbidden states $\mathcal{M}_B$ and the first column, the set of the simplified over-states $B_3$.

| $c_j \downarrow \quad M_i \rightarrow$ | $P_1P_4P_6$ | $P_2P_4P_6$ | $P_2P_3P_7$ | $P_2P_4P_7$ | $P_2P_5P_7$ |
|---|---|---|---|---|---|
| $P_2P_4$ | 0 | 1 | 0 | 1 | 0 |
| $P_2P_5$ | 0 | 0 | 0 | 0 | 1 |
| $P_4P_6$ | 1 | 1 | 0 | 0 | 0 |
| $P_2P_7$ | 0 | 0 | 1 | 1 | 1 |
| $Cv(M_i)$ | 1 | 2 | 1 | 2 | 2 |

Fig. 3. Function $R(c_j, M_i)$ and $Cv(M_i)$

By observing this table, we conclude that $\forall M_i \in \mathcal{M}_B$ $Cv(M_i) \geq 1$, and thus the set of non forbidden states $\mathcal{M}_E$ is equal to the set of authorized states $\mathcal{M}_A$.

## VII. IMPLEMENTATION OF CONTROLLER

### A. Final covering

After the simplifications presented above, it is possible to choose the simplest constraints covering all the forbidden states. In the result of the last step, the same forbidden state can be covered by several over-states. The rules to choose the final over-states are similar with the rules of the Quine-McCluskey method to simplify the logical expressions [13]. To choose the final results, the table of Figure 3 is used and modified in Figure 6.

| $c_j \downarrow \quad M_i \rightarrow$ | $P_1P_4P_6$ | $P_2P_4P_6$ | $P_2P_3P_7$ | $P_2P_4P_7$ | $P_2P_5P_7$ | $B_4$ |
|---|---|---|---|---|---|---|
| $P_2P_4$ | 0 | **1** | 0 | **1** | 0 | - |
| $P_2P_5$ | 0 | 0 | 0 | 0 | **1** | - |
| $P_4P_6$ | **1** | **1** | 0 | 0 | 0 | 1 |
| $P_2P_7$ | 0 | 0 | **1** | **1** | **1** | 1 |
| $Cf(M_i)$ | **1** | **1** | **1** | **1** | **1** | |

Fig. 6. Function $R(c_j, M_i)$ and $Cf(M_i)$

To choose the *minimal set of constraints*, denoted as $B_4$, firstly it is necessary to choose the over-state such as the forbidden state can be covered only by this one ($Cv(M_i)=1$), when this one exists. Then we mark all the forbidden states, which correspond to it in the line $Cf(M_i)$. This line corresponds to the final covering. If a forbidden state is covered by two or several over-states, it is necessary to choose the over-state which covers the most non selected forbidden states. In the case of equality, the simplest over-state will be selected.

**Corollary 1:** The set of the non forbidden states $\mathcal{M}_E$ defined by the set of the constraints deduced from $B_4$ is equal to $M_A$ if and only if: $\quad \forall M_i \in \mathcal{M}_B \quad Cf(M_i) = 1$

❑

This corollary means that it is necessary for each forbidden states to be covered at least by one over-state. When this is verified, the maximal permissive controller is obtained.

### B. Controller synthesis

The set of the constraints equivalent to $B_4$ is denoted as $C_4$. To calculate the control places corresponding to each linear constraint, we will use the method developed in [4]. This technique will be called *invariant approach* recalled briefly below. Let $W_R$ be the incidence matrix of the system (process and specifications). Each place of the controller will add a line to this matrix. Let $W_{RC}$ be the incidence matrix of the PN model corresponding to the controlled system. It is made up of two matrices, the original matrix of system $W_R$ and the incidence matrix of the controller $W_C$. From the set of constraints $C_4$, the matrix $L$ and the constant vector $C\_bound$ can be constructed. It is possible to calculate in an algebraic way the incidence matrix of the controller as it is presented below. $M_{Ri}$ is the initial marking of system $R$ and $M_{Ci}$ is the control places initial marking. The very simple way to calculate $W_C$ makes this approach very popular.

$$W_C = -L.W_R, \quad M_{C\_i} = C\_bond - L.M_{R\_i}$$

Let us take again the example of Figure 1, the set of final constraints ($C_4$) is:

$$m_4 + m_6 \leq 1, \quad m_2 + m_7 \leq 1$$

$$L = \begin{bmatrix} 0 & 0 & 0 & 1 & 0 & 1 & 0 \\ 0 & 1 & 0 & 0 & 0 & 0 & 1 \end{bmatrix}$$

$$W_C = -L.W_R$$

$$\Rightarrow W_C = \begin{bmatrix} 0 & 1 & -1 & 1 & -1 \\ -1 & 0 & 0 & 0 & 1 \end{bmatrix}$$

$$M_{C\_i} = C\_bond - L.M_{R\_i}$$

$$\Rightarrow M_{c1} = 0; \quad M_{c2} = 1$$

Yamalidou [4] showed that if all the events are controllable, the controller is maximal permissive. However, if there are uncontrollable events, the extended method presented in [3] does not give the optimal solution in the general case. The problem exists when a control place is synchronized with a place of the process by an uncontrollable event. In this case, the process cannot respect the PN firing rules. It means that the set of the reachable states will be greater than that the set given by the PN model. We will indicate this set as $\mathcal{A}_{RC}$.

We are going to show that if Corollary 1 is true, the controller obtained is maximal permissive even if uncontrollable transitions exist.

**Remark 4:** A marking of the set $\mathcal{A}_{RC}$ differs from a marking of $\mathcal{M}_E$ because of the added control places. This is only a coding of these sets. To be able to compare the various sets of states, we will omit the control places for the elements of the set $\mathcal{A}_{RC}$.

❑

**Property 4:** Let $\mathcal{M}_E$ be the set of authorized states by the constraints deduced from $B_4$ and let $\mathcal{A}_{RC}$ be the automaton corresponding to the set of accessible state in the controlled system, If $\mathcal{M}_E = \mathcal{M}_A$ then $\mathcal{A}_{RC}$ is isomorphic to $\mathcal{M}_E$ and the controller obtained by the invariant approach is maximal permissive.

❑

In the case of our example the function $Cf(M_i)$ (final covering ) is equal to 1 for each $M_i \in \mathcal{M}_B$, therefore $\mathcal{M}_E = \mathcal{M}_A$ (Corollary 1) then the controller is maximal permissive (Property 4). The PN model of the final controller is represented in Figure 8.

It should be noticed that there are some control places with output uncontrollable transitions. However, that never leads to a bad behavior, i.e. when a control place is not marked; there is at least one non marked input place of this uncontrollable transition, which belongs to the process. Moreover, controllable events $c_1$ and $c_2$ have been removed since the control is now performed by the control places. The different sets computations are in polynomial complexity except for the $\mathcal{M}_B$ over-states computation which is exponential. Fortunately the number of border sates is often small.

## VIII. CONCLUSION

In this paper, we have presented a systematic method to reduce the number of linear constraints corresponding to the forbidden states for a safe PN. This is realized by using the non reachable states and by building the constraints with a systematic method. The important concept of over-state has been defined; it corresponds to a set of markings having the same property (forbidden or authorized). From the forbidden states the set of over-states is calculated. The utilization of over-state concept allows simplifying the constraints. Properties giving necessary and sufficient conditions for the existence of a maximal permissive controller were established. After the simplifications, the existence of the controller is proved formally. When this one exists, the invariant approach allows computing the controller.

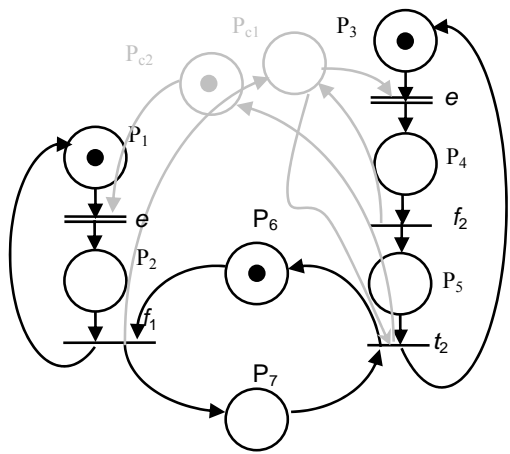

Fig. 8. PN Model in closed loop with control places